\documentclass[a4paper]{spie}  

\usepackage{amsmath,amsfonts,amssymb}
\usepackage{graphicx}
\usepackage[colorlinks=true, allcolors=blue]{hyperref}

\title{LOCNES: Low Cost NIR Extended Solar Telescope}

\author[a]{Claudi R.}
\author[b]{Ghedina A.}
\author[c]{Pace E.}
\author[c]{Gallorini L.}
\author[d]{Di Giorgio A.--M.}
\author[d]{Liu S.--J.}
\author[e]{Tozzi A.}
\author[a]{Carleo I.}
\author[f]{Lanza A.F.}
\author[g]{Micela G.}
\author[h]{Molinari E.}
\author[b]{Poretti E.}
\author[g]{Phillips D.}
\author[i]{Tripodo G.}

\affil[a]{INAF Astronomical Observatory of Padova, vicolo Osservatorio, 5, Padova, Italy}
\affil[b]{INAF -- Fundaci\'on Galileo Galilei, Rambla Jos\'e Ana Fern\'andez P\'erez, 7,  Bre$\tilde{n}$a Baja--TF, Spain}
\affil[c]{Dep. of Physics and Astronomy, Universit\'a degli studi di Firenze, Firenze, Italy}
\affil[d]{INAF-- IAPS, via del Fosso del Cavaliere, 100, Roma, Italy}
\affil[e]{INAF--Astrophysical Observatory of Arcetri, Largo Enrico Fermi,5, Firenze, Italy}
\affil[f]{INAF--Astrophysical Observatory of Catania, Via S.Sofia 78, Catania, Italy}
\affil[g]{INAF--Astronomical Observatory of Palermo,P.zza del parlamento, 1, Palermo, Italy}
\affil[h]{INAF--Astronomical Observatory of Cagliari,Via della Scienza, 5, 09047 Cuccuru Angius, Selargius (CA), Italy}
\affil[i]{Universit\'a di Palermo, Scuola delle Scienze di base e applicate, Dip. di Fisica e Chimica, Piazza Marina, 61, 90133, Palermo (Italy)}

\authorinfo{Further author information: (Send correspondence to R.Claudi)\\R. Claudi: E-mail: riccardo.claudi@inaf.it}

\begin{document} 
\maketitle

\begin{abstract}
The search for telluric extrasolar planets with the Radial Velocity (RV) technique is intrinsically limited by the stellar jitter due to the activity of the star, because stellar surface inhomogeneities, including spots, plages and convective granules, induce perturbations hiding or even mimicking the planetary signal. This kind of noise is poorly understood in all the stars, but the Sun, due to their unresolved surfaces. For these reasons, the effects of the surface inhomogeneities on the measurement of the RV are very difficult to characterize. On the other hand, a better knowledge of these phenomena can allow us a step forward in our understanding of solar and stellar RV noise sources. This will allow to develop more tools for an optimal activity correction leading to more precise stellar RVs. Due to the high spatial resolution with which the Sun is observed, this noise is well known for it. Despite this, a link is lacking between the single observed photospheric phenomena and the behavior of the Sun observed as a star. LOCNES (Low Cost NIR Extended Solar Telescope) will allow to gather time series of RVs in order to disentangle the different contributions to the stellar (i.e., sun’s) RV jitter.  Since July 2015, a Low Cost Solar Telescope (LCST) has been installed outside the TNG dome to feed solar light to the HARPS-N spectrograph (0.38-0.69 $\mu$m; R=115000). The refurbishment of the Near Infrared (NIR) High Resolution Spectrograph GIANO (now GIANO-B) and the new observing mode GIARPS at TNG (simultaneous observations in visible with HARPS-N and in NIR with GIANO-B) is a unique opportunity to extend the wavelength range up to 2.4 $\mu$m for measuring the RV time series of the Sun as a star. This paper outlines the LOCNES project and its scientific drivers.  
\end{abstract}


\section{INTRODUCTION}
\label{sec:intro}  
The search for very small extrasolar planets with the Radial Velocity technique is plagued by the stellar jitter due to the activity of the star, because stellar surface inhomogeneities including spots, plages and granules, induce perturbation that hide the planetary signal. This kind of noise is poorly understood in all the stars but the Sun due to their unresolved surfaces. For these reasons the effect of the surface inhomogeneities on the measurement of the Radial velocities are very difficult to characterize. On the other hand, a better understanding of these phenomena can allow us to made a step further in our knowledge of solar and stellar physics. The latter will allow to acquire more skills in the art of developing optimal correction techniques to extract true stellar radial velocities. A viable way to tackle these problems is to observe the Sun as a star getting time series of radial velocities in order to disentangle the several contribution of the stellar (solar) RV jitter\cite{dumusqueetal2015,marchwinskietal2015,haywoodetal2016}. 

With the aim to obtain long-term observation of the Sun as a star with state-of-the-art sensitivity to RV changes, since July 2015, a Low Cost Solar Telescope (LCST) has been installed\cite{phillipsetal2016} on the outside of the dome at TNG to feed solar light to the HARPS-N spectrograph (0.38-0.69um; R=115000). The main purpose of the LCST is to show that HARPS-N with the Astro Laser Frequency Comb calibrator\cite{philipsetal2012} have enough precision and stability to measure on the spectra of the Sun the Center of Mass (CoM) Radial Velocity (RV; of the order of 9cm/s) due to the gravitational pull of an earth-like planet (Venus for the Sun).
While the LCST and HARPS-N continue in an autonomous way to acquire spectra, the interest on the collected data is growing day after day. Thanks to the resolved images of the surface of the Sun produced by the SORCE and SDO satellites (together with magnetograms and doppler maps) it is possible to correlate the observed spectra with surfaces anomalies like spots, plages/flares, granulation etc.

In the last decade, new near -- infrared (NIR) spectrograph have been built that are able to measure high precision RVs and thanks to the advent of better technology, specifically in calibration, these spectrographs are able to approach the precision that are routinely obtained in the visible. Among these NIR spectrograph there is the high resolution NIR spectrograph of the Telescopio Nazionale Galileo (TNG) named GIANO--B. The Italian exoplanetary community through the {\it Progetto Premiale WOW} funded the refurbishment of GIANO-A\cite{olivaetal2012} in the framework of the GIARPS project\cite{claudietal2017}. The Near Infrared (NIR) GIANO spectrograph (0.9-2.5um, R=50000) of the TNG underwent a refurbishment improving the throughput and the efficiency. The spectrograph was moved from the Nasmyth A focus of the TNG, where it was fed by optical fibers, to the Nasmyth B focus where it is fed by a preslit optics. The main results are: enhancement of the efficiency of the spectrograph; the elimination of the modal noise due to injection of K band in the Z-Blan fibers and the added advantage of placing GIANO-B close to HARPS--N. The use of a dichroic optical system allows the simultaneous collection of a VIS-NIR spectrum of the observed target. This observing mode, called GIARPS, is a worldwide unique and exclusive facility of the TNG. 
Taking advantage of the presence of GIARPS we plan to replicate the opto-mechanical configuration of the LCST but extending the wavelength range to the NIR in order to simultaneously feed both HARPS--N and GIANO--B.
The VIS part is fundamental in order to be able to observe simultaneously with the LCST and LOCNES Telescope (feeding the 2 fibers of HARPS--N\footnote{The second fiber that will feed HARPS--N is already under discussion with the HARPS--N and LCST utilizer. In any case we consider it as well because it is part of the LOCNES project}) and check that the same spectra are produced on HARPS--N.  The NIR part will extend to the full band-width of GIANO-B (0.9 to 2.5um).

\section{Science drivers}
\label{sec:sciencedrivers}  
High Resolution spectrographs are used for high precision radial velocity measurements in the search for extrasolar planets evaluating the Doppler shift of the spectral lines of the star due to the presence of a low mass companion. In this task, these instruments become more and more precise also because people apply the two techniques of the simultaneous thorium and absorbing cell. The precision reached such a low limit (the best precisions obtained nowadays are of the order of 0.3 m/s rms) 
that the detection of very small planets (or planets in wider orbits) has to tackle with also low level of star activity. In fact, beside the instrumental noise the radial velocity technique is also hampered by astrophysical sources of noise mainly due to the host star itself like, for example, the stellar pulsations, surface granulation and stellar jitter caused by star spots or other instabilities in the stellar atmosphere. At the 1 m/s level of precision, in fact, these physical phenomena in stellar photospheres give significant signals that can hide planetary radial-velocity signatures if not properly modeled. Solar-type stars have an outer convective envelope that exhibits variability on different timescales. There are several sources of noise that are necessary to take into account in high precision radial velocity measurements. In any case the signal of a planet and that due to activity should be distinguishable because of the different physical origin. But, so far, this is not the case. 
Understanding the RV signatures of stellar activity, especially those at the stellar rotation time-scale, is essential to better understand the short and very short time scale variability of stars, but also to improve our ability to detect and characterize (super-)Earths and even small Neptunes in orbits of a few days to weeks. 

All the sources of these stellar noise are inherent to stellar surface inhomogeneity, but, up to now it was only possible to obtain precise, frequent, full- disk RV measurements derived from the full optical spectrum for stars other than the Sun. However, the surfaces of other stars are not resolvable; and therefore all the information used to study surface inhomogeneities is indirect, including: photometric flux (e.g., \citenum{boisseetal2009,aigrainetal2012}), bisector of spectral lines or of the cross correlation function depending on the activity level of the star (e.g., \citenum{vogt1987,quelozetal2001,figueiraetal2013,dumusqueetal2014}), or the calcium chromospheric activity index \cite{noyesetal1984}. Because it is extremely difficult to infer from such indirect observables the size, location, and contrast of surface inhomogeneities, understanding in detail the RV perturbations produced by these features is a major challenge. 
This is true in particular today that RV searches started to target young and active stars. 

The Sun is the only star whose surface can be directly resolved at high resolution, and therefore constitutes an excellent test case to explore the physical origin of stellar radial -- velocity (RV) variability. The contemporaneous observation of the Sun as a star in both the VIS and in the NIR will give  us the possibility to better understand the effect of surface inhomogeneities on RVs by obtaining precise full-disk RV measurements of the Sun. This approach of observing the Sun as a star allows us to directly correlate any change in surface inhomogeneities observed by solar satellites like the Solar Dynamics Observatory (SDO \cite{pesnelletal2012}) with variations in the full-disk RV.

   \begin{figure} [ht]
   \begin{center}
   \begin{tabular}{c} 
   \includegraphics[height=5cm]{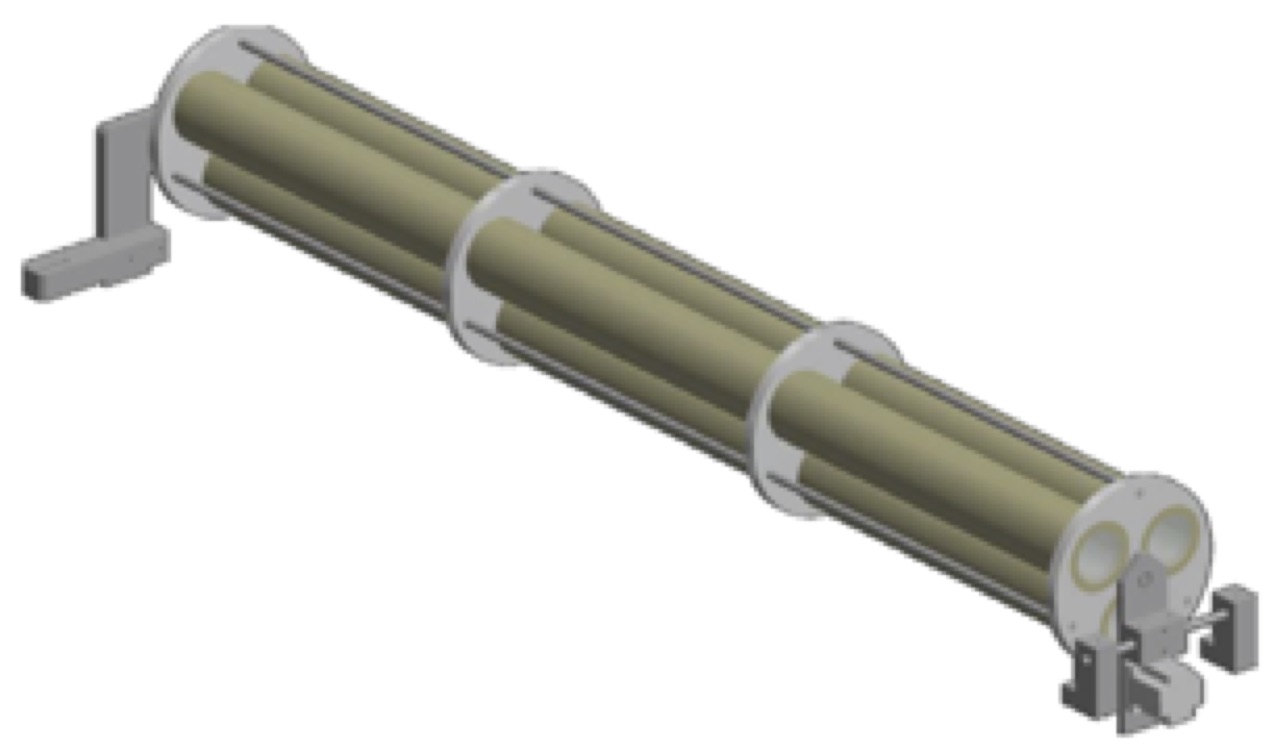}
   \end{tabular}
   \end{center}
   \caption[example] 
   { \label{fig:celle} 
Mechanical design for the GIANO cell revolver. The revolver mechanism mounts two cells (and an open slot), and rotates into the optical beam. The revolver is located in the preslit area of GIANO-B \cite{tozzietal2016} in a dedicated structure. }
   \end{figure} 

\subsection{The K Band Justification}
\label{ssec:kband}  
The measurements of the sun-as-a-star radial velocity was performed by Deming \& Plymate\cite{demingandplymate1994} by means of the molecular bands of CO around 2.3 microns, thus the K band is of fundamental importance to measure the Sun-as-a-star radial velocity variations in the infrared domain. In the case of distant stars, this passband has been used to measure the RV and correct the variations induced by stellar activity, e.g., by means of the spectrograph CSHELL at NASA IRTF or NIRSPEC@Keck \cite{crockettetal2012}. Therefore, access to the K band is required to compare solar and stellar measurements.
Observation of the disc-integrated solar spectrum in the infrared K passband are crucial for an understanding of the effects of solar magnetic activity on the variations of the solar radial velocity. This happens because the K band samples layers of the solar atmosphere that are different from those that are sampled by the J and H passbands. Specifically, in the level sampled by the K passband, the brightness distribution, the velocity fields, and the magnetic field intensity are different than in those of the J and H passbands, allowing us a complete mapping of the photospheric convection and its perturbations in active regions (cf. \citenum{penn2014}). These results can be used to understand how magnetic fields affect the radial velocity measurements of the Sun as a star. Only in the Sun, we can compare disk-integrated measurements with spatially resolved maps of the active regions to test our models of the radial velocity perturbations to be used to correct activity effects in distant stars. The availability of data in the K passband is crucial to improve those models. 
Moreover, the Ti line at 2231 nm is an excellent probe of the relatively cooler plasma in sunspots and can be used to measure their effects on the disc-integrated flux, velocity fields, and mean magnetic field (e.g., \citenum{pennetal2003}).

\subsection{Sun's Radial Velocity measurements in the NIR Range}
GIANO-B will be equipped with NIR absorbing cells in order to have an inertial inner reference for high precision radial velocity measurements. The cells will be mounted on a revolver inside the preslit optics (Figure \ref{fig:celle}) to be able to put the absorbing cell into the optical path towards the GIANO--B slit.

In this way, a spectrum of CH$_4$-C$_2$H$_2$-NH$_3$, the gas mixture present into the absorbing cell \cite{seemanetal2018}, is over imposed to the stellar spectrum. The value of the radial velocity of the star is so evaluated by fit of the composite spectrum by a synthetic reconstruction of it using the cell spectrum, the instrumental profile of GIANO-B and a high signal to noise spectrum of the star alone. The absorbing cell technique is described in Butler et al (\citenum{butleretal1996}). 
In this moment, the cells are under construction and we developed an alternative procedure to get the value of RVs by the telluric lines.
The method to extract radial velocities (RV) follows the approach described in Carleo et al (\citenum{carleoetal2016}): the telluric lines are used as wavelength reference and the cross-correlation function (CCF) method is used to determine the stellar RV. For this purpose, we start with the preparation of files including evaluation of the correction to the barycenter of the solar system and a re--sampling of spectra in order to have a constant step of 200 m/s in RV. Then the spectra are normalized by dividing them to a fiducial continuum. One of the most important steps for NIR RVs is the removal of the telluric lines from science spectra. For this purpose, we create the best telluric spectrum from a library including several telluric spectra acquired over time. This best telluric spectrum is shifted to take into account small shifts in the dispersion zero points, and scaled in intensity to take into account the different airmass between telluric and science spectra. Finally, the obtained telluric spectrum is subtracted from the science spectra. This approach, that determines the best shift and scale parameters, allows to minimize the RMS of the subtracted spectra. 
From the best telluric spectrum and the cleaned stellar spectra, we construct the digital telluric and stellar masks respectively. Then, the procedure performs the cross correlation of individual orders of the normalized spectra with these masks (both stellar and telluric), with derivation of individual CCF and RVs. Through the weighted sum of the individual CCFs we obtain the stellar and the telluric RVs: the latter are finally subtracted from the former, providing the final relative stellar RVs. The uncertainties are then evaluated taking into account the photon statistics. As a final step, we derive the bisector velocity span (BIS) of the CCF. The whole procedure requires about 1.5 minutes per spectrum. 
Anyway, a slightly different approach can be considered (Carleo et al. 2018 submitted): it takes into account the weight of the single orders, being some of them affected by telluric lines and thus contributing in different ways to the RVs determination. So, starting from the RV of individual orders RVo, we calculate the scatter among the exposures $\sigma_0$, and the corresponding weight, w$_0$. The weighted RV for each order RV$_{o,wo} =$RV$_o$/V$_o$ is then used to evaluate the weighted RV for each exposure:$\sum{RV_o}/\sum{w_0}$. The corresponding error is $1/\sqrt{\sum{w_0}}$.

   \begin{figure} [ht]
   \begin{center}
   \begin{tabular}{c} 
   \includegraphics[height=8cm]{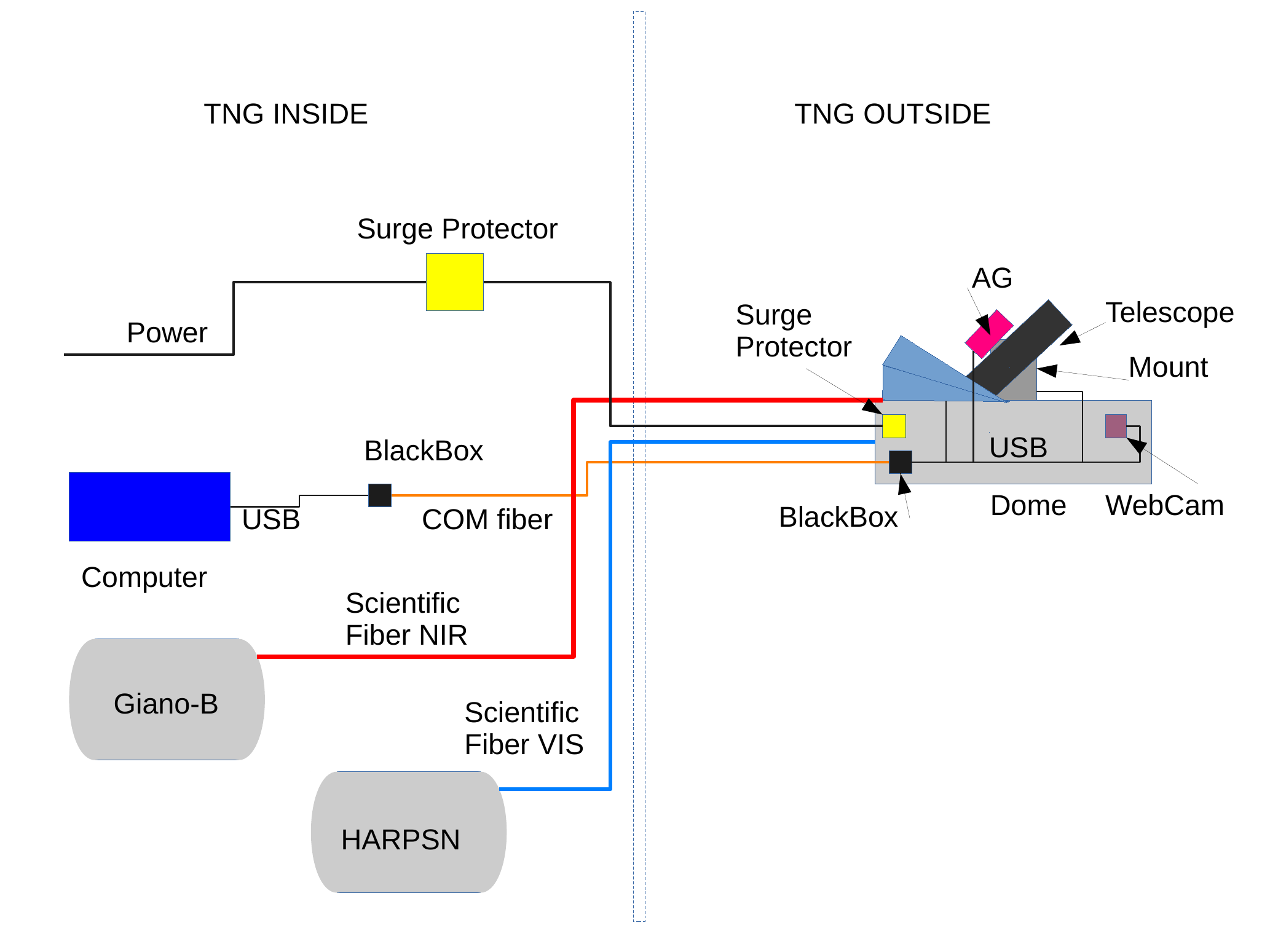}
   \end{tabular}
   \end{center}
   \caption[example] 
   { \label{fig:scheme} 
LOCNES system scheme. All main part are indicated together with the connections to the control PC and spectrographs inside the TNG structure.}
   \end{figure} 

\section{LOCNES}
\label{sec:locnes}
All the LOCNES system is described Figure\ \ref{fig:scheme} where besides the telescope there are also the other main parts of LOCNES. Into the dome (a custom made one) in addition to the telescope there are all the electronics need for the control of the dome and the mount of the teslescope and for the communication with the GIANO--B and HARPS--N spectrographs. Anti--static filters and a rain sensor are also hosted by the dome. The telescope will be mounted on a platform fastened on the external stairs flight of the TNG dome. Two patches of optical fiber, one to feed HARPS--N and the second to feed GIANO-B, depart from the dome towards the inside of the TNG, with a path of about 40 m, down to Nasmyth--B platform where the two spectrographs are located. In the following the main parts of LOCNES telescope are briefly described.

\subsection{The telescope}
\label{ssec:tel}
\begin{figure} [ht]
   \begin{center}
   \begin{tabular}{c} 
   \includegraphics[height=7cm]{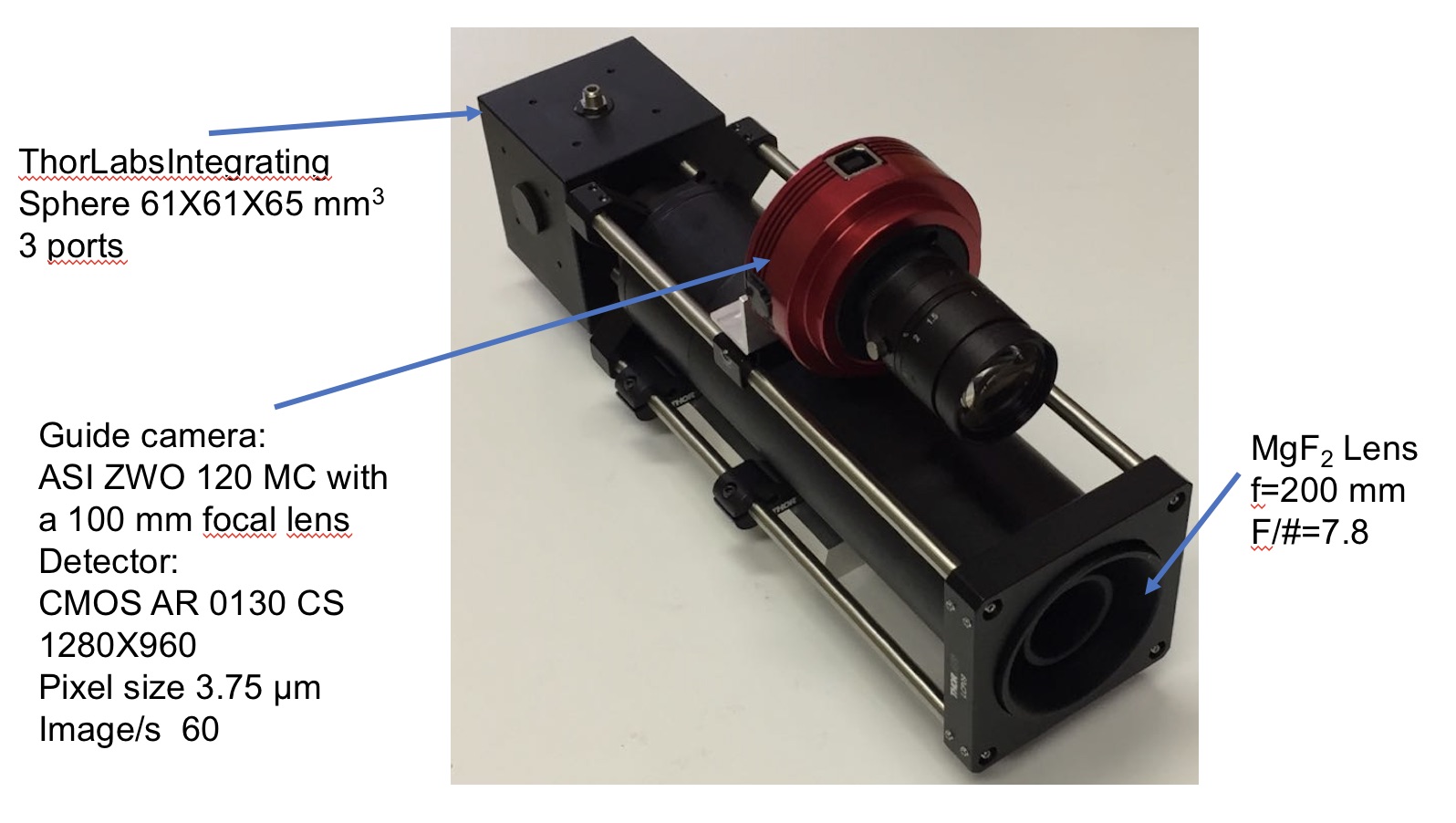}
   \end{tabular}
   \end{center}
   \caption[example] 
   { \label{fig:telescope} 
 The LOCNES telescope.}
   \end{figure}

The body of the LOCNES Telescope (see Figure\ \ref{fig:telescope}) is constituted by four main parts: the primary aperture, the structure, the mount and the integrating sphere (IS). From the integrating sphere, 2 fibers depart towards the two spectrographs (HARPS--N and GIANO--B). On the structure of the telescope is mounted a CMOS--camera with the aim of guiding. 

\begin{figure} [ht]
   \begin{center}
   \begin{tabular}{c} 
   \includegraphics[height=5cm]{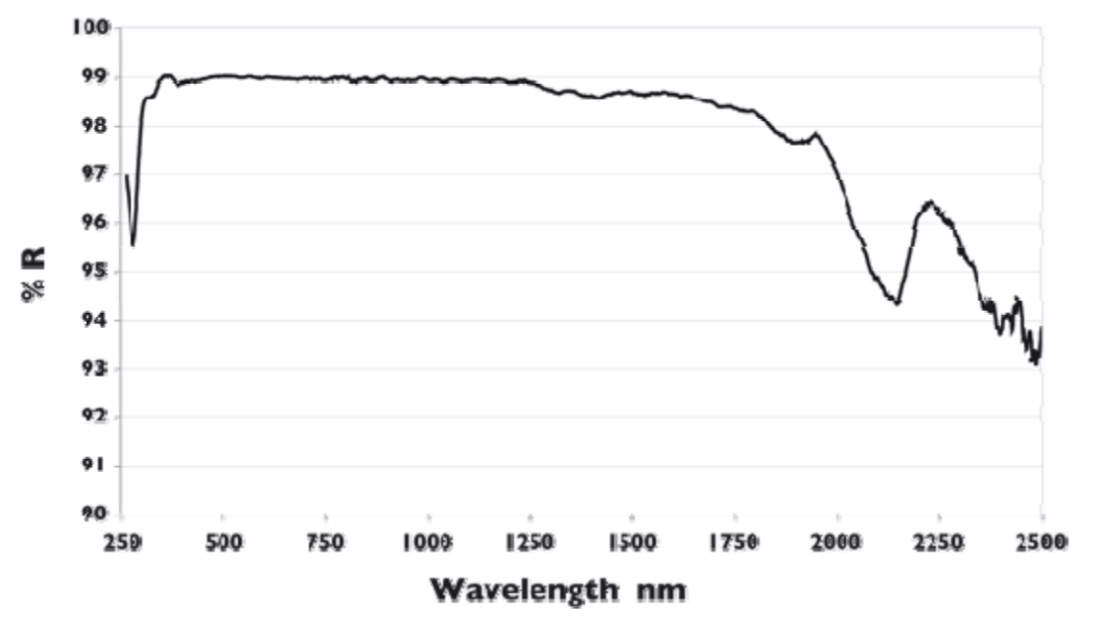}
   \end{tabular}
   \end{center}
   \caption[example] 
   { \label{fig:is} 
 The LOCNES Integrating Sphere Reflectivity (reproduced by Thorlabs IS200 data sheet \footnote{https://www.thorlabs.com/}).}
   \end{figure}

The aperture of the telescope is constituted by a lens mounted on a commercial holder. The lens is a MgF$_2$ plano -- covex lens with a diameter of 25.4 mm and a focal distance of f=200 mm. The lens feeds the integrating sphere that is positioned at the focal distance of the lens. The IS is a Thor Labs IS200 series $61 \times 61 \times 65$ mm$^3$ with four ports. The IS has a reflectance of about 99\% in the wavelength range between 350 and 1500 nm and $ >95\%$ in the remaining part of the whole wavelength range between 250 and 2500 nm (see Figure\ \ref{fig:is}). The lens and the integrating sphere are connected by the telescope structure, all built with commercial pieces by Thorlabs. 

The structure of the telescope is then connected to a commercial AZ motorized mounts that assure the tracking of the Sun. The mount selected is a iOptron AZ\footnote{http://www.ioptron.com/product-p/8900.htm}. 

\subsection{The dome}
\label{ssec:dome}
\begin{figure} [hb]
   \begin{center}
   \begin{tabular}{c} 
   \includegraphics[height=7cm]{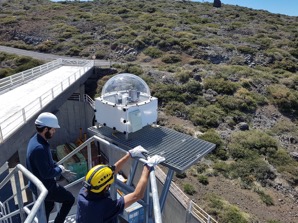}
   \end{tabular}
   \end{center}
   \caption[example] 
   { \label{fig:loc} 
 The platform outside the TNG where LOCNES will be mounted. In the figure is visible the already existing LCST.}
   \end{figure}

To allow simple, unattended operation, the telescope will be placed under a fixed dome located near to the preexisting LCST (see Figure\ \ref{fig:loc}). All control signals will be sent from the control computer located inside the TNG to the LOCNES via USB over fiber as well as from the solar telescope to the GIANO-B via fiber to minimize electrical connections between LOCNES and the TNG. 

\begin{figure} [ht]
   \begin{center}
   \begin{tabular}{c} 
   \includegraphics[height=7cm]{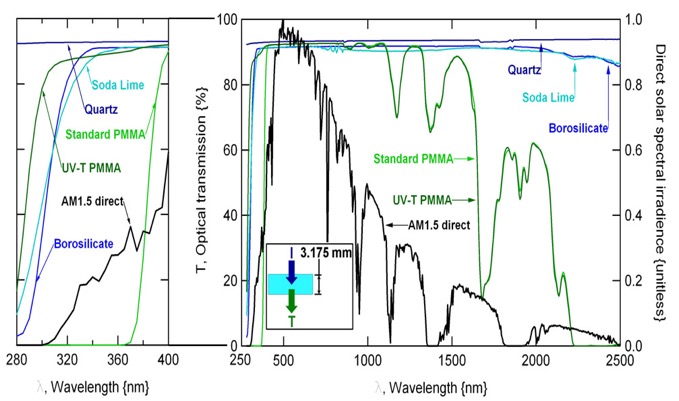}
   \end{tabular}
   \end{center}
   \caption[example] 
   { \label{fig:PMMA} 
 The transmission of several possible materials for the construction of a fixed and transparent dome compared with the Sun irradiation. The LCST dome is made in standard PMMA that is not transparent to the NIR.}
   \end{figure}

The dome will be different from that of LCST because the acrylic solution cut or reduce a lot the transmission in the NIR part of the wavelength range (see Figure\ \ref{fig:PMMA}). The other possible material with which made a fixed and transparent dome are too expensive. We choose to have a custom made dome in aluminum. The dome is built in a way that the aperture of petals is towards the North with an obscuration in this direction of about 30° up the horizon. The South field of view is completely free from east to west.
The closure of the dome is water and dust proof. The opening and closing of the dome is made by a CC motor. The external diameter of the dome is 580 mm (500 mm inner diameter) and is posed on a sealed, stainless steel box of 400 mm of height and a base of $600\times600$\ mm$^2$ that is the site of the power and control electronics. From the basement of the LOCNES dome all the cables and the fiber are confined inside a flexible plastic conduit fixed to the LOCNES box by a pass through.

\subsection{The science fibers}
\label{ssec:fib}
\begin{figure} [bt]
   \begin{center}
   \begin{tabular}{c} 
   \includegraphics[height=6cm]{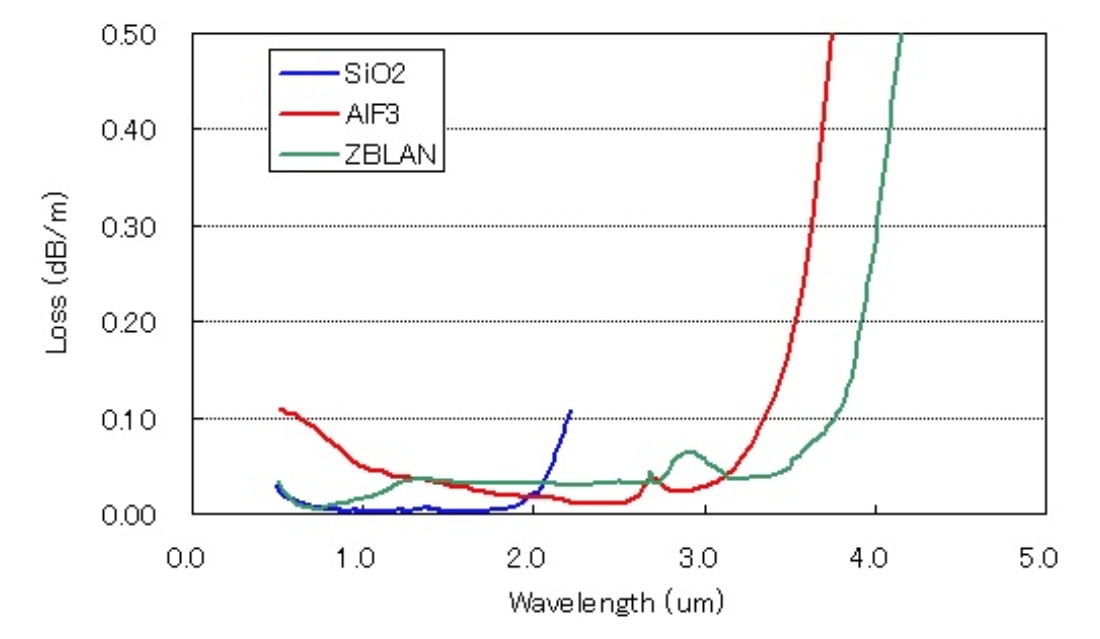}
   \end{tabular}
   \end{center}
   \caption[example] 
   { \label{fig:fiber} 
 The attenuation of available NIR Optical Fibers. The green line is the attenuation of ZBLAN fibers.}
   \end{figure}

As described in Section\ \ref{ssec:kband}, the K--band of the Sun's flux is important for the scientific aims we want to achieve with LOCNES. This implies that the 40 m long optical fiber patch shall attenuate as low as possible the whole NIR wavelength range (up to K band) accepted by GIANO--B. To fulfill the requirement we have to choose the Z--BLAN (ZrF$_4$/BaF$_2$/LaF$_3$/AlF$_2$/NaF) fluoride fibers that transmit from the visible up to $4\ \mu$m (see Figure\ \ref{fig:fiber}).

Due to their brittleness, that causes a very little tolerance to the mechanical stresses, like small radius bending, and high costs, we use these fibers (ZBLAN) just for the GIANO--B feeding. As for HARPS--N, the injection of the Sun's light from fibers into the GIANO--B spectrograph is done with the connection of the fibers with the spectrograph's calibration unit. The transfer of Sun's light from LOCNES to the HARPS--N will be instead performed with the same type of fiber which is currently feeding HARPS--N from LCST.

\section{The Schedule}
\label{ssec:schedule}
\begin{figure} [ht]
   \begin{center}
   \begin{tabular}{c} 
   \includegraphics[height=9cm]{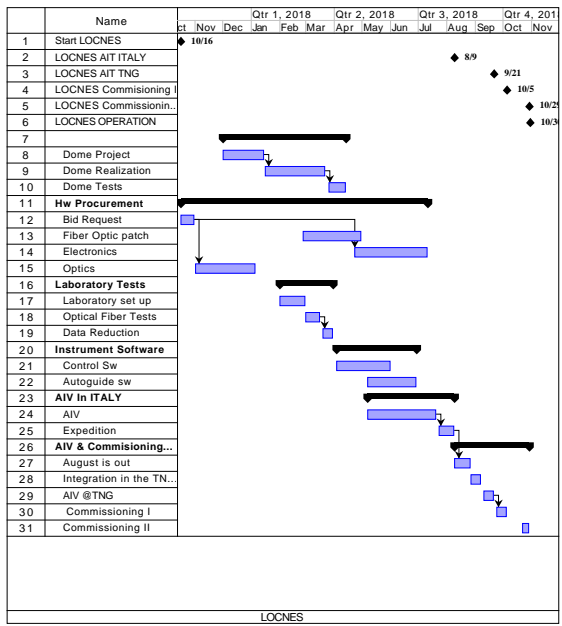}
   \end{tabular}
   \end{center}
   \caption[example] 
   { \label{fig:sche} 
 The time schedule of the LOCNES Project}
   \end{figure}

The development of the telescope will last about 1 yr and after this time lapse, LOCNES will be in operation.
Figure\ \ref{fig:sche} shows the schedule of the project. 

The main milestones are the following:
\begin{enumerate}
\item 2017 – October – 16: kick off meeting of LOCNES
\item 2018 – June – 27: end of the AIV in Italy and shipping of LOCNES to the TNG
\item 2018 – September – 14: end of AIV phase at the TNG
\item 2018 – October – 5: end of the first commissioning phase (Integration with the telescope and spectrographs control system)
\item 2018 – October – 29: end of the second commissioning phase
\item 2018 – October – 30: start of Science observations.
\end{enumerate}

\section{Conclusions}
\label{sec:conclusion}

We outlined the scientific aims and the description of the LOCNES telescope project. LOCNES telescope will begin its duty in Fall 2018, recording, thanks to the two high resolution spectrographs of the TNG, VIS and NIR time series of Sun's radial velocity measurements. The wide wavelength range of LOCNES will help to connect the stellar (solar) activity causes and the different zones and depths of the chromosphere to the resulting RVs allowing the development of more tools for an optimal activity correction. Furthermore, the TNG acquires a low cost simple instrument that makes it the unique astronomical structure in the world with the possibility to gather high resolution spectra of the Sun as a star  from the visible up to the K band.

\acknowledgments 
 
 Authors would like to thank the support by INAF through Progetti Premiali funding scheme of the Italian Ministry of Education, University, and Research.

\bibliography{report} 
\bibliographystyle{spiebib} 

\end{document}